# On Control Schemes of Voltage Source Converters

Fahmid Sadeque and Tareq Hossen

*Abstract*— This paper discusses some aspects of control schemes for voltage source converters under abnormal conditions. The control schemes are developed specifically for the situations when one or more system parameters vary significantly to the extent that the system becomes unstable with a conventional controller. The paper will present some recent works on control of grid-interactive converters for parameter variations in weak grid. The paper will also discuss some methods under abnormal dc-bus variation. Finally, the paper focuses on the control schemes suitable for ac-dc converters that addresses input frequency, voltage and load variation. All the discussed control schemes have shown robust performance under abnormal conditions.

## I. INTRODUCTION

Voltage source converters are gradually becoming integrated parts of our life. Application of the voltage source converters can span from grid-interactive inverters, to household appliances, to versatile forms of electric vehicles [1]-[4]. The grid-interactive inverters enable high integration of renewable energy sources while also allowing for remote and dynamic control [5]-[10]. Many states in North America have integrated renewable energy resources in their power system planning for the future. Los Angles, California aims to achieve 100% of renewable electricity by 2045 along with aggressive electrification targets for buildings and vehicles [11]. Furthermore, California's total solar power generation is nearly 13%, with certain places generating as much as 25% [12]. In [13], it is reported that California has set a goal of generating 50% of its energy from renewable sources by 2030. This increasing penetration of DGs allows for greater flexibility in power networks, where grid-interactive voltage source converters are the most important part of the modern power grid [14]-[17]. Likewise, the transportation system has already started moving toward electrification of technologies, i.e., more electrification instead of mechanical controls, replacement/modification of power-trains, replacement of combustion engines [18]-[19], etc. With modern power-electronic technologies, the ultimate goal is to obtain high fuel efficiency, low emission, and low maintenance costs. Therefore, this technological transformation can be achieved through the voltage source converters. Notice, this massive integration of voltage source converters in different aspects has brought up many challenges, which have been and are being solved through many different types of control scheme. This paper specifically focuses on the recent trend in research in the control scheme of voltage source converters under abnormal conditions, i.e., parameter variation in internal and external abnormalities, dc-bus voltage variation, weak and unbalanced grid, rapid frequency, voltage, and load variation, etc. The following paragraphs will focus on the abnormal conditions in different categories of application of voltage source converters.

The widespread use of DGs has many advantages, but it also poses new issues in terms of power system stability and reliability [20]-[23]. An inertia-less inverter-based DG results in low-inertia microgrids. Specifically, in a weak grid, an inverter could become unstable and, therefore, would have to be disconnected from the system [24]. Weak grid results from large grid impedance seen at the point of common coupling (PCC). When the PCC voltage contains harmonic components at the filter's natural frequency, instability may occur [25]. The variation of weak-grid's grid-impedance can cause instability from unwanted resonances [26]. Even more, the voltage feedforward path used in conventional voltage source controller for smaller response time can also cause instability under grid-impedance variation. Furthermore, the grid-distortion can affect with controller in weak grid [27]. Moreover, the phase-angle estimation capability of the phase-locked loop also deteriorates under unbalanced and distorted grid conditions [28].

Inverters should have the ability to identify the internal and the external abnormal conditions and must have the control techniques to operate effectively [29]-[30]. The conventional two-level three-phase dc/ac voltage source converter offers less opportunities for fault tolerant control in comparison to the multi-level three-phase inverters. Specifically, the cascaded H-bridge (CHB) multi-level inverter is mostly preferred for high-power, medium voltage operations because of its inherent fault tolerant capability. On the other hand, the performance of the CHB can be negatively impacted by the variation of dc-bus voltage and hence impacting the conventional PWM methods [31]. In addition, while injecting active and reactive power to the grid, the voltage converters may incur dc bus variation. This may result fluctuation in power injection with conventional PWM reference. Under this condition, the PWM reference should be adjusted so that, the available DC bus utilization is maximized [32]-[35].

The voltage source converters in electric vehicles presents new challenges to the researchers. One of the major challenges is to obtain the capability to adapt for rapid input-frequency/voltage variations, load fluctuations, and extreme ambient conditions changes while regulating the desired parameters with seamless dynamics and satisfactory power quality by the voltage source converters. In literature, different power converter control schemes have been developed for more electric power-trains. The controller needs to be fast and accurate irrespective of the system parameter. Hence, the controller needs to be adaptive also [36], [37].

It is evident from the discussion of the previous paragraphs, that the capability of the recent controllers for the voltage source converters should not be limited to the conventional output signal generation. Rather, the controllers should be adaptive, fast, and accurate to overcome different adverse scenarios. In this respect, this paper focuses on some state-of-the-art control for voltage source converters that addresses the aforementioned challenges. Apart from the introduction, the rest of the paper is organized as follows. Section II discusses on some advanced control strategies under weak-grid conditions. Section III discusses on the corrective schemes of multi-level inverters and control strategies under dc-bus variation. Section IV focuses on voltage source controller under rapid frequency and load variation. Finally, section V concludes the paper with a discussion of future research scopes.

## II. Voltage Source Controller for Weak Grid Parameter Variation

This section discusses on some advanced control methods for voltage source converters under weak grid scenarios. Several advanced techniques are available to ensure the stable operation in weak grid. The most common approach includes modified feedforward paths in the controller [18], [38]-[40]. The authors in [39] introduced a capacitor voltage feedforward method for improving the adaptability of the voltage converter. There, a delay compensation link has been added to the voltage feedforward path that reduces the interaction between the grid-distortion and the control scheme. An impedance-phase compensation strategy is proposed in [40]. In [41], PCC voltage feedforward is employed through a filter and a gain block. It is shown that for the modified voltage feedforward, with a suitable value of the gain block, the converter exhibits improved performance in weak grid. It is mention worthy that, all these methods have a tradeoff between the steady-state performance of the closed-loop grid-tied system.

In [42], a virtual inductance feedforward control scheme is developed to enhance the stability of grid-tied VSIs in weak grids. A virtual inductance term is derived emulating the impact of the additional grid-side filter without adding large filer inductors. In this method, the measured current is fed to the inner current control loops through a gain block, known as the virtual inductance. Fig. 1 shows a controller equipped with virtual inductance for stability enhancement of a grid-interactive inverter under weak grid. Notice, the controller does not require any additional measurements/sensors. Notice, despite the feedforward technique in [42] demonstrates improved performance in weak grids, it requires manual adjustment of the feedforward term. Implementation of the techniques may become difficult if the grid-impedance of the system is unknown. Hence, in [43] and [44], two methods have been introduced where the gain parameters. The adaptive techniques use the same principles as the previously described. In addition, the gain parameters of the feedforward paths are updated adaptively. In [45] a direct model reference adaptive method (MRAC) is integrated to the virtual inductance feedforward scheme to adaptively vary the virtual inductance for changes in grid impedance in weak grids. The direct MRAC method is integrated to the PQ controller to develop a modified PQ controller with adaptive virtual inductance feedforward. In [45], an adaptive control scheme to enhance the stability of inverters is presented based on the online estimation of grid impedance. For larger values of grid impedance, the PLL bandwidth was lowered to keep the inverter in the stable region. However, the PLL bandwidth had to be lowered considerably to ensure stability, thus introducing a tradeoff between stability and dynamic performance. In [45], an active damper is added to the system, which essentially introduces an additional resistive term in the inverter circuit that can be varied adaptively to make the inverter more robust against changes in grid impedance. Among the control methods discussed above, the methods derived in [43] is the only method that does not require a parameter estimation stage and can adapt the virtual inductance value according to a stable reference model and hence, is more robust. A detailed derivation and analysis of the method in [43] is also available in [46]-[48].

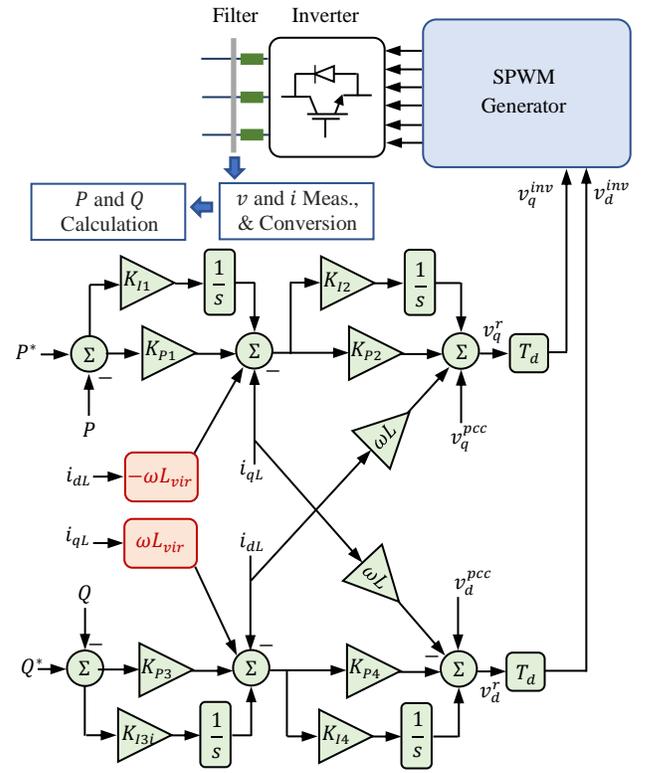

Fig. 1. Stability enhancement of weak-grid through introduction of virtual inductance in the controller.

## III. Voltage Source Controller for Time-Variant DC Sources

In this section, some corrective scheme-based control for voltage source converters are presented. Specifically, this paper section is focused on the CHB controller capable to operate under dc-bus variation under abnormal conditions. As the conventional multi-level PWM techniques are unable to effectively utilize the time-variant DC sources with unequal voltage magnitudes in CHB, many applications ensure the equal magnitudes of the sources. In [49] a double star connected converter, and in [50], a boost inverter (see [51], [52] for details) topology has been presented. These methods do not directly address the time-variant source problems and hence, can make the topology more complex. The time variant nature of the dc sources can be resolved through dc-dc converter topologies to regulate the input dc voltage [49]-[50], [53]. Although this provides a solution to the problem, the additional circuitry introduces extra weight, cost, size and loss to the whole system. Another solution is to implement the active balancing techniques [54], [55], which either requires additional components and control structure, or connection between multiple dc sources, eliminating the isolation between sources required for CHB converters. To compensate for the dc-bus oscillation, an approach is to add common-mode component to the line-to-neutral output voltages of three-phase three-wire voltage source converters without affecting the converters' current [56], [57]. A number of techniques for creating appropriate common-mode components have been developed and are suitable for implementation in a wide range of converter topologies [58], [59]. A well-known method is to inject the third-harmonic components. In [60], the common-mode components to ensure linear-modulation is determined as the mean value of a set of common mode elements. However, this method only considers implementation in balanced system.

In [32] and [33] an atypical PWM technique has been presented to compensate for the dc-bus voltage variation for CHB converters. The common-mode component injected into PWM references is based on the available dc bus voltages and enables references to be adjusted in real-time, without requiring lookup tables. In [2] a similar atypical PWM technique has been presented to voltage source converter that can provide symmetrical and asymmetrical ancillary services during dc-bus oscillation. The technique provides harmonic compensation as symmetrical ancillary services, and negative sequence compensation while providing asymmetrical ancillary services. Notice, for both techniques, the PWM signals are modulated as per the dc-bus fluctuation, so that the fluctuation is compensated. Fig. 2 shows the block diagram utilizing the negative sequence controller for grid-following inverters.

## IV. Voltage Source Controller for Rapid Input Frequency variation

The voltage source converter may undergo rapid input frequency variation. Specifically, the ac/dc converters used in the electric vehicles, i.e., more electric aircrafts, where the mechanical energy conversion system has been partially, or completely replaced by the voltage source converters. One of the major requirements of these voltage source converters is to be applied for variable-speed operations. This variable-speed operation needs to generate a wide-range variable-frequency/ constant-amplitude or a variable-frequency/amplitude set of voltages, which constitutes an entirely novel operation scenario for power converters that are typically implemented at constant industrial frequency levels in grid-tied [38], [61]-[63], stand-alone applications [30] or narrower variable frequency ranges in motor drives or renewable energy interfacing [64]. In following paragraph, the recent voltage source control schemes for variable input frequency variation has been highlighted.

The voltage source converters for electric vehicles can be of different types based on their tasks, i.e., DC/AC. AC/AC. DC/DC. and AC/DC power converters [19], [36]-[37]. The voltage converter used for motor drive control the propulsion motor and hence, need to operate bi-directionally to utilize the advantage of regenerative braking [65]. Some fault-tolerant motor drives have been extensively discussed in [66]-[68]. These methods are robust under symmetric conditions. However, with any variation in input frequency these controllers will not be helpful. Some direct and indirect matrix converters can be used as ac/ac converters to be utilized in-between for frequency variation [69], [70]. Nevertheless, this add extra circuitry, weight, and size in the system, which is not expected for electric vehicles.

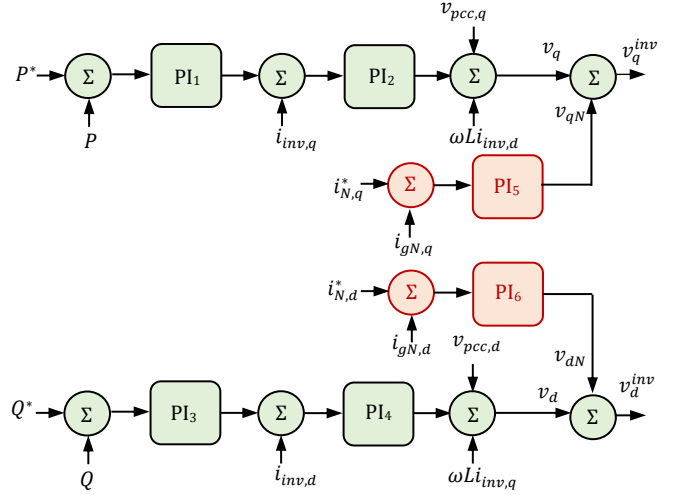

Fig. 2. Utilization of negative-sequence controller for grid-following inverter.

In [29], a step-ahead model predictive controller for has been proposed to perform under input frequency variation. This controller can exhibit ultrafast dynamics in dc-bus and power factor regulation when the converter is operated for variable-frequency/constant-amplitude input voltage. Its fast dynamics has been achieved from the combination of the model predictive control and instantaneous phase-angle detection technique. However, the model predictive control schemes may suffer from performance degradation under model parameter mismatch and parameter variation [75-76]. The control method in [36] may become erroneous as it requires accurate parameter values to correctly switching the duty ratios. Hence, in [37], an adaptive filter parameter estimation technique is adopted in addition to the contribution of [36]. Herein, the dynamic performance can be improved by adopting Lyapunov-based adaptive parameter estimation algorithm, which can provide accurate-fast tracking of the parameters of the system and then fed to the step-ahead controller. The only drawback of these types of estimations is, it introduces steady-state tracking error in input reactive power. The drawbacks have been resolved in [71] and [72], where a direct model reference adaptive controller (MRAC) is presented. This new controller is capable of seamlessly regulate the output dc-bus voltage and input reactive power during ultrafast frequency/voltage variations. In this technique, the controller's gains are automatically adjusted to the external perturbations regardless of parameter values of the system. Fig. 3 depicts a block diagram of the controller presented in [72].

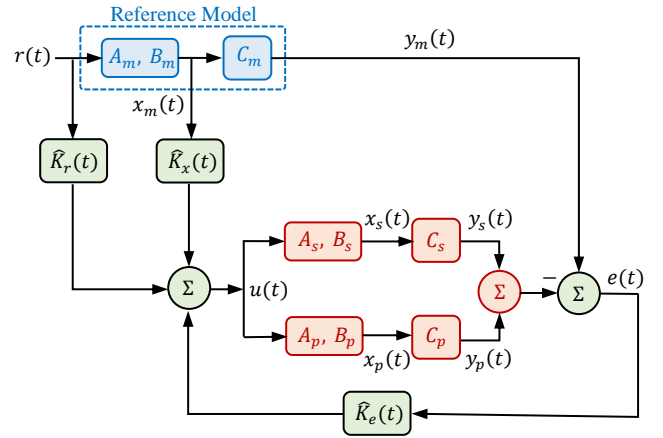

Fig. 3. Block diagram of a direct model reference adaptive controller.

## V. Conclusion

In this article some recent control schemes for different types voltage source converters performing under variable input/output parameters have been discussed. Specifically, the paper highlighted the contributions on robustness of the controller under system parameter variation in weak grid, input dc-bus voltage variation in unbalanced grid, and input frequency variation in the voltage source converters in the more electric vehicles. A direct model reference adaptive method (MRAC) has been found effective under grid-parameter variation during weak grid condition. The controller can adaptively estimate the virtual inductance based on the change in the grid impedance in a weak grid and add the estimation to the feedforward path to eventually ensure a stable control. Then, an atypical PWM technique has been highlighted, that effectively compensate for the dc-bus voltage variation for CHB converters. The technique can provide harmonic compensation as symmetrical ancillary services, and negative sequence compensation while providing asymmetrical ancillary services. Finally, a direct model reference adaptive controller (MRAC) under wild frequency variation has been discussed. The controller can automatically adjust the gains to the external perturbations regardless of parameter values of the system. All the aforementioned voltage source controllers have shown impressive robust performance under one or more parameter fluctuation scenarios and have potential to be extended to any applications in power grid, motor drive, and electric vehicle technologies.

## VI. References

[1] B. Mirafzal and A. Adib, "On grid-interactive smart inverters: features and advancements," *IEEE Access*, vol. 8, pp. 160526-160536, 2020.
[2] A. Adib, J. Lamb and B. Mirafzal, "Ancillary services via VSIs in microgrids with maximum DC-bus voltage utilization," *IEEE Trans. Ind. Appl.*, vol. 55, no. 1, pp. 648-658, Jan.-Feb. 2019.
[3] A. Adib, K. K. Afridi, M. Amirabadi, F. Fateh, M. Ferdowsi, B. Lehman, L. H. Lewis, B. Mirafzal, M. Saeedifard, M. B. Shadmand, and P. Shamsi, "E-mobility - advancements and challenges," *IEEE Access*, vol. 7, pp. 165 226–165 240, 2019.
[4] J. Benzaquen, F. Fateh and B. Mirafzal, "On the dynamic performance of variable-frequency AC–DC converters," *IEEE Trans. Transport. Electrific.*, vol. 6, no. 2, pp. 530-539, June 2020.
[5] S. K. Mazumder *et al.*, "A Review of Current Research Trends in Power-Electronic Innovations in Cyber-Physical Systems," in *IEEE J. Emerg. Sel. Topics Power Electron*.
[6] I. Serban, S. Céspedes, C. Marinescu, C. A. Azurdia-Meza, J. S. Gomez, and D. S. Hueichapan, "Communication requirements in microgrids: A practical survey," *IEEE Access*, vol. 8, pp. 47694–47712, 2020.
[7] O. Dag and B. Mirafzal, "On stability of islanded low-inertia microgrids," *2016 Clemson University Power Systems Conference (PSC)*, 2016, pp. 1-7.
[8] B. Mirafzal, M. Saghaleini, and A. K. Kaviani, "An SVPWM-based switching pattern for stand-alone and grid-connected three-phase single-stage boost-inverters," *IEEE Trans. Power Electron.*, vol. 26, no. 4, pp. 1102 - 1111, April 2011.
[9] A. Singh, A. K. Kaviani, and B. Mirafzal, "On dynamic models and stability analysis of three-phase phasor PWM-based CSI for stand-alone applications," *IEEE Trans. Ind. Electron.*, vol. 62, no. 5, pp. 2698 – 2707, May 2015.
[10] A. Singh and B. Mirafzal, "Three-phase single-stage boost inverter for direct drive wind turbines," *2016 IEEE Energy Conversion Congress and Exposition (ECCE)*, Milwaukee, WI, 2016, pp. 1-7.
[11] C. Jaquelin, and P. Denholm, "The Los Angeles 100% renewable energy study," National Renewable Energy Laboratory, Golden, CO, USA, NREL/TP-6A20-79444. Available: https://maps.nrel. gov/la100/.
[12] D. J. Sebastian and A. Hahn, "Exploring emerging cybersecurity risks from network-connected DER devices," *2017 North American Power Symposium (NAPS)*, Morgantown, WV, 2017, pp. 1-6.
[13] J. Qi, A. Hahn, X. Lu, J. Wang and C. Liu, "Cybersecurity for distributed energy resources and smart inverters," *IET Cyber-Physical Syst.: Theory Appl.*, vol. 1, no. 1, pp. 28-39, 12 2016.
[14] M. Gursoy and B. Mirafzal, "On self-security of grid-interactive smart inverters," *2021 IEEE Kansas Power and Energy Conference (KPEC)*, 2021, pp. 1-6.
[15] H. Han, X. Hou, J. Yang, J. Wu, M. Su, and J. M. Guerrero, "Review of power sharing control strategies for islanding operation of AC microgrids," *IEEE Trans. Smart Grid*, vol. 7, no. 1, pp. 200–215, Jan. 2016.
[16] M. S. Pilehvar and B. Mirafzal, "PV-fed smart inverters for mitigation of voltage and frequency fluctuations in islanded microgrids," 2020 *International Conference on Smart Grids and Energy Systems (SGES)*, 2020, pp. 807-812.
[17] M. S. Pilehvar, M. B. Shadmand and B. Mirafzal, "Analysis of smart loads in nanogrids," *IEEE Access*, vol. 7, pp. 548-562, 2019.
[18] J. Benzaquen, J. He and B. Mirafzal, "Toward more electric power-trains in aircraft: Technical challenges and advancements," *CES Tran. Electr. Mach. Syst.*, vol. 5, no. 3, pp. 177-193, Sept. 2021.
[19] J. Benzaquen, F. Fateh, M. B. Shadmand and B. Mirafzal, "Performance comparison of active rectifier control schemes in more electric aircraft applications," *IEEE Trans. Transport. Electrific.*, vol. 5, no. 4, pp. 1470-1479, Dec. 2019.
[20] F. Blaabjerg, Z. Chen, and S. B. Kjaer, "Power electronics as efficient interface in dispersed power generation systems," *IEEE Trans. Power Electron.*, vol. 19, no. 4, pp. 1184–1194, Sep. 2004.
[21] J. Yin, S. Duan and B. Liu, "Stability analysis of grid-connected inverter with LCL filter adopting a digital single-loop controller with inherent damping characteristic,"*IEEE Trans. Ind. Informatics*, vol. 9, no. 2, pp. 1104-1112, May 2013.
[22] J. Rocabert, A. Luna, F. Blaabjerg, and P. Rodriguez, "Control of power converters in AC microgrids," *IEEE Trans. Power Electron.*, vol. 27, no. 11, pp. 4734–4749, Nov. 2012.
[23] A. K. Kaviani, and B. Mirafzal, "Stability analysis of the three-phase single-stage boost inverter," *IEEE Applied Power Electronics Conference & Exposition (APEC)*, March 2013, pp. 918-923.
[24] F. Blaabjerg, R. Teodorescu, M. Liserre, and A. V. Timbus, "Overview of control and grid synchronization for distributed power generation systems," *IEEE Trans. Ind. Electron.*, vol. 53, no. 5, pp. 1398–1409, Oct. 2006.
[25] C. Zheng, L. Zhou, X. Yu, B. Li, and J. Liu, "Online phase margin compensation strategy for a grid-tied inverter to improve its robustness to grid impedance variation," *IET Power Electron.*, vol. 9, no. 4, pp. 611-620, Mar. 2016.
[26] M. Lu, A. Al-Durra, S. M. Muyeen, S. Leng, P. C. Loh, and F. Blaabjerg, "Benchmarking of stability and robustness against grid impedance variation for LCL-filtered grid-interfacing inverters," *IEEE Trans. Power Electron.*, vol. 33, no. 10, pp. 9033-9046, Oct. 2018.
[27] Y. Huang, X. Yuan, J. Hu and P. Zhou, "Modeling of VSC Connected to Weak Grid for Stability Analysis of DC-Link Voltage Control," *IEEE J. Emerg. Sel. Topics Power Electrons.*, vol. 3, no. 4, pp. 1193-1204, Dec. 2015.
[28] F. Sadeque, J. Benzaquen, A. Adib and B. Mirafzal, "Direct phase-angle detection for three-phase inverters in asymmetrical power grids," *IEEE J. Emerg. Sel. Topics Power Electron.*, vol. 9, no. 1, pp. 520-528, Feb. 2021.
[29] B. Mirafzal, "Survey of fault-tolerance techniques for three-phase voltage source inverters," *IEEE Trans. Ind. Electron.*, vol. 61, no. 10, pp. 5192-5202, Oct. 2014.
[30] J. Lamb, and B. Mirafzal, "Open-circuit IGBT fault detection and location isolation for cascaded multilevel converters," *IEEE Transactions on Industrial Electronics,* vol. 64, no.6, pp. 4846 - 4856, June 2017.

[31] J. Lamb and B. Mirafzal, "Grid-interactive cascaded h-bridge multilevel converter PQ plane operating region analysis," *IEEE Trans. Ind. Appl.*, vol. 53, no. 6, pp. 5744-5752, Nov.-Dec. 2017.

[32] J. Lamb, B. Mirafzal and F. Blaabjerg, "PWM common mode reference generation for maximizing the linear modulation region of CHB converters in islanded microgrids," *IEEE Trans. Ind. Electron.*, vol. 65, no. 7, pp. 5250-5259, July 2018.

[33] J. Lamb and B. Mirafzal, "An adaptive SPWM technique for cascaded multilevel converters with time-variant DC sources," *IEEE Trans. Ind. Appl.*, vol. 52, no. 5, pp. 4146-4155, Sept.-Oct. 2016.

[34] J. Lamb and B. Mirafzal, "Active and reactive power operational region for grid-interactive cascaded h-bridge multilevel converters," *2016 IEEE Energy Conversion Congress and Exposition (ECCE)*, Milwaukee, WI, USA, 2016, pp. 1-6.

[35] A. Singh and B. Mirafzal, "An efficient grid-connected three-phase single-stage boost current source inverter," *IEEE Power Energy Technol. Syst. J.*, vol. 6, no. 3, pp. 142-151, Sept. 2019.

[36] J. Benzaquen, A. Adib, F. Fateh and B. Mirafzal, "A model predictive control scheme formulation for active rectifiers with LCL filter," in *Proc. 2019 IEEE Energy Conversion Congress and Exposition (ECCE)*, 2019, pp. 3758-3763.

[37] J. Benzaquen, F. Fateh, M. B. Shadmand and B. Mirafzal, "One-step-ahead adaptive control scheme for active rectifiers in wild frequency applications," in *Proc. 2019 IEEE Applied Power Electronics Conference and Exposition (APEC)*, 2019, pp. 588-593.

[38] A. Adib, B. Mirafzal, X. Wang and F. Blaabjerg, "On stability of voltage source inverters in weak grids," *IEEE Access*, vol. 6, pp. 4427-4439, 2018.

[39] V. S. Pour-Mehr, B. Mirafzal, and O. Mohammed, "Pulse-load effects on ship power system stability," *IEEE Industrial Electronics Society Conf.*, Nov. 2010, pp. 3353-3358.

[40] X. Li, J. Fang, Y. Tang, X. Wu, and Y. Geng, "Capacitor-voltage feedforward with full delay compensation to improve weak grids adaptability of LCL-filtered grid-connected converters for distributed generation systems," *IEEE Trans. Power Electron.*, vol. 33, no. 1, pp. 749-764, Jan. 2018.

[41] G. Wang, X. Du, Y. Shi, Y. Yang, P. Sun, and G. Li, "Effects on oscillation mechanism and design of grid-voltage feedforward in grid-tied converter under weak grid," *IET Power Electron.*, vol. 12, no. 5, pp. 1094-1101, May 2019.

[42] A. Adib and B. Mirafzal, "Virtual inductance for stable operation of grid-interactive voltage source inverters," *IEEE Trans. Ind. Electron.*, vol. 66, no. 8, pp. 6002-6011, Aug. 2019.

[43] A. Adib, F. Fateh and B. Mirafzal, "Smart inverter stability enhancement in weak grids using adaptive virtual-inductance," *IEEE Trans. Ind. Appl.*, vol. 57, no. 1, pp. 814-823, Jan.-Feb. 2021.

[44] J. Xu, S. Xie, Q. Qian, and B. Zhang, "Adaptive feedforward algorithm without grid impedance estimation for inverters to suppress grid current instabilities and harmonics due to grid impedance and grid voltage distortion," *IEEE Trans. Ind. Electron.*, vol. 64, no. 9, pp. 7574-7586, Sep. 2017.

[45] L. Jia, X. Ruan, W. Zhao, Z. Lin, and X. Wang, "An adaptive active damper for improving the stability of grid-connected inverters under weak grid," *IEEE Trans. Power Electron.*, vol. 33, no. 11, pp. 9561-9574, Nov. 2018.

[46] A. Adib, F. Fateh, M. B. Shadmand, and B. Mirafzal, "Weak grid impacts on stability of voltage source inverters - Asymmetrical Grid," *IEEE Energy Conversion Congress and Exposition (ECCE),* September 2018.

[47] A. Adib, F. Fateh, M. B. Shadmand, and B. Mirafzal, "A reduced-order technique for stability investigation of voltage source inverters," *IEEE Energy Conversion Congress and Exposition (ECCE),* September 2018.

[48] A. Adib, F. Fateh, B. Mirafzal, "Weak grid impacts on the design of voltage source inverters" *IEEE Workshop on Control and Modeling for Power Electronics,* June 2018.

[49] M. Quraan, T. Yeo and P. Tricoli, "Design and control of modular multilevel converters for battery electric vehicles," *IEEE Trans. Power Electron.*, vol. 31, no. 1, pp. 507-517, Jan. 2016.

[50] Z. Du, B. Ozpineci, L. M. Tolbert and J. N. Chiasson, "DC–AC cascaded h-bridge multilevel boost inverter with no inductors for electric/hybrid electric vehicle applications," *IEEE Trans. Ind. Appl.*, vol. 45, no. 3, pp. 963-970, May-June 2009.

[51] A. Singh, and B. Mirafzal, "Indirect boost matrix converter and low-voltage generator for direct drive wind turbines," *J. Eng.,* vol. 2018, no. 1, pp. 10-16, 1 2018.

[52] Singh, A. A. Milani, and B. Mirafzal, "Modified phasor pulse width modulation method in three phase single stage boost inverter," *IEEE Applied Power Electronics Conference & Exposition (APEC)*, March 2014, pp. 1276-1280.

[53] W. Qian, H. Cha, F. Z. Peng and L. M. Tolbert, "55-kw variable 3x dc-dc converter for plug-in hybrid electric vehicles," *IEEE Trans. Power Electron.*, vol. 27, no. 4, pp. 1668-1678, April 2012.

[54] C. Young, N. Chu, L. Chen, Y. Hsiao and C. Li, "A single-phase multilevel inverter with battery balancing," *IEEE Trans. Ind. Electron.*, vol. 60, no. 5, pp. 1972-1978, May 2013.

[55] H. Park, C. Kim, K. Park, G. Moon and J. Lee, "Design of a charge equalizer based on battery modularization," *IEEE Trans. Veh. Technol.*, vol. 58, no. 7, pp. 3216-3223, Sept. 2009.

[56] B. Mirafzal, "Survey of fault-tolerance techniques for three-phase voltage source inverters," *IEEE Trans. Ind. Electron.*, vol. 61, no. 10, pp. 5192-5202, Oct. 2014.

[57] D. S. Ochs, B. Mirafzal, and P. Sotoodeh, "A method of seamless transitions between grid-tied and stand-alone modes of operation for utility-interactive three-phase inverters," *IEEE Trans. Ind. Appl.*, vol. 50, no. 3, pp. 1934–1941, May/June 2014.

[58] W. Yao, H. Hu and Z. Lu, "Comparisons of space-vector modulation and carrier-based modulation of multilevel inverter," *IEEE Trans. Power Electron.*, vol. 23, no. 1, pp. 45-51, Jan. 2008.

[59] M. Aleenejad, H. Mahmoudi and R. Ahmadi, "Unbalanced space vector modulation with fundamental phase shift compensation for faulty multilevel converters," *IEEE Trans. Power Electron.*, vol. 31, no. 10, pp. 7224-7233, Oct. 2016.

[60] F. Carnielutti, H. Pinheiro and C. Rech, "Generalized carrier-based modulation strategy for cascaded multilevel converters operating under fault conditions," *IEEE Trans. Ind. Electron.*, vol. 59, no. 2, pp. 679-689, Feb. 2012.

[61] M. Gursoy and B. Mirafzal, "Self-security for grid-interactive smart inverters using steady-state rference model," in Proc. 2021 IEEE 22st Workshop on Control and Modeling for Power Electronics (COMPEL).

[62] A. Adib, F. Fateh and B. Mirafzal, "A stabilizer for inverters operating in grid-feeding, grid-supporting and grid-forming modes," in *Proc.* 2019 IEEE Energy Conversion Congress and Exposition (ECCE), 2019, pp. 2239-2244.

[63] M. S. Pilehvar and B. Mirafzal, "A frequency control method for islanded microgrids using energy storage systems," in *Proc.* 2020 IEEE Applied Power Electronics Conference and Exposition (APEC), 2020, pp. 2327-2332.

[64] A. Singh, J. Benzaquen and B. Mirafzal, "Current source generator–converter topology for direct-drive wind turbines," *IEEE Trans. Ind. Appl.*, vol. 54, no. 2, pp. 1663-1670, March-April 2018.

[65] A. K. Kaviani, B. Hadley and B. Mirafzal, "A time-coordination approach for regenerative energy saving in multiaxis motor-drive systems," *IEEE Trans. Power Electron.*, vol. 27, no. 2, pp. 931-941, Feb. 2012.

[66] A. Sayed-Ahmed, B. Mirafzal and N. A. O. Demerdash, "Fault-tolerant technique for Δ-connected AC-motor drives," *IEEE Trans. Energy Convers.*, vol. 26, no. 2, pp. 646-653, June 2011.

[67] S. Choi et al., "Fault diagnosis techniques for permanent magnet AC machine and drives—A review of current state of the art," *IEEE Trans. Transport. Electrific.*, vol. 4, no. 2, pp. 444-463, June 2018.

[68] B. Mirafzal, R. J. Povinelli and N. A. O. Demerdash, "Interturn Fault Diagnosis in Induction Motors Using the Pendulous Oscillation Phenomenon," *IEEE Trans. Energy Convers.*, vol. 21, no. 4, pp. 871-882, Dec. 2006.

[69] P. Wheeler et al., "A matrix converter based permanent magnet motor drive for an aircraft actuation system," in *Proc.* IEEE International Electric Machines and Drives Conference, 2003. IEMDC'03., 2003, pp. 1295-1300 vol.2.

[70] J. Lei, B. Zhou, J. Wei, X. Qin and J. Bian, "Aircraft starter/generator system based on indirect matrix converter," in *Proc.* IECON 2014 - 40th Annual Conference of the IEEE Industrial Electronics Society, 2014, pp. 4840-4846.

[71] J. Benzaquen and B. Mirafzal, "Seamless dynamics for wild-frequency active rectifiers in more electric aircraft," *IEEE Trans. Ind. Electron.*, vol. 67, no. 9, pp. 7135-7145, Sept. 2020.

[72] J. Benzaquen, M. B. Shadmand and B. Mirafzal, "Ultrafast rectifier for variable-frequency applications," *IEEE Access*, vol. 7, pp. 9903-9911, 2019.